\documentclass[twocolumn,showpacs,amsmath,amssymb]{revtex4}
\input{epsf}

\usepackage{graphicx}
\usepackage{array}
\usepackage{bigstrut}
\usepackage{longtable}
\usepackage{rotating,booktabs}
\usepackage{booktabs,threeparttable}
\usepackage{bm}
\usepackage{lipsum}
\usepackage{color}
\usepackage{mathtools}

\begin{document}
	
\title{Role of negative-energy states on the E2-M1 polarizability of optical clocks}


\author{Fang-Fei Wu, Ting-Yun Shi, Wei-Tou Ni, and Li-Yan Tang$^{\dag}$~\footnotetext{\dag Email Address: lytang@apm.ac.cn}}

\affiliation {State Key Laboratory of Magnetic Resonance and Atomic and Molecular Physics, Wuhan Institute of Physics and
Mathematics, Innovation Academy for Precision Measurement Science and Technology, Chinese Academy of Sciences, Wuhan 430071, People's Republic of China}

\date{\today}

\begin{abstract}
The theoretical calculations of the dynamic E2-M1 polarizability at the magic wavelength of the Sr optical clock are inconsistent with experimental results. We investigate role of negative-energy states in the E2 and M1 polarizabilities. Our result for E2-M1 polarizability difference $-$7.74(3.92)$\times$10$^{-5}$ a.u. is dominated by the contribution from negative-energy states to M1 polarizability and has the same sign as and consistent with all the experimental values. In addition, we apply the present calculations to various other optical clocks, further confirming the importance of negative-energy states to the M1 polarizability. 
\end{abstract}

\maketitle

\emph{Introduction.} Optical clocks have advanced to an unprecedented level of stability, precision, and sensitivity~\cite{brewer19a,sanner19a,roberts2020,lange2021,beloy2021,bothwell22a}. An expected realization in the redefinition of frequency and time using optical clocks will be in the near future~\cite{bregolin17a,yamanaka15a,ludlow15a}. Optical clocks are being used to test Einstein equivalence principle and to search for variations of constant~\cite{godun14a,huntemann14a,safronova18a,bothwell22a}. Further improvement would enable the implementation of new scenes, such as in the space detection of gravitational waves with AU-sized network~\cite{kolkowitz16a,ebisuzaki20a,ni16a}.

Both optical lattice clocks and optical ion clocks show significant contributions from the Stark shift due to thermal radiation to the total clock uncertainty~\cite{nicholson15a,ushijima15a,mcgrew18a,brewer19a,bothwell19a,oelker19a,lu22a,huang22a}. The accurate determination and theoretical understanding of the Stark shift is crucial for the improvement of optical clocks. For an atom in a laser field, the energy levels shift due to the frequency-dependent multipolar polarizabilities of the atomic states~\cite{manakov86book}. To cancel the dominant electric dipole (E1) Stark shift of the transition, the optical clock is working at the magic wavelength~\cite{takamoto05a,ludlow06a}. However, when the precision of optical clocks is reaching $10^{-18}$ or beyond, the contributions of electric quadrupole (E2) and magnetic dipole
(M1) polarizabilities become significant~\cite{ovsiannikov13a,katori15a,porsev18a,ushijima18a,westergaard11a}. 

\begin{figure}
\includegraphics[width=0.52\textwidth]{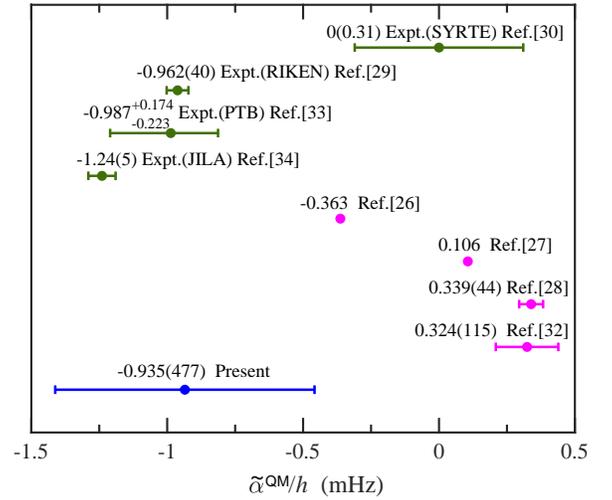}
\caption{\label{f1}(Color online) Comparison of the E2-M1 polarizability difference  $\tilde{\alpha}^{QM}/h$ (in mHz) for the Sr clock. Values in green are experimental results, values in magenta denotes other theoretical results, while the value marked in blue is from our present calculation.}
\end{figure}

For the Sr optical clock, the E2-M1 polarizability difference at the magic wavelength of 813.4280(5) nm~\cite{ye09a} between theory~\cite{porsev18a, wu2019a,ovsiannikov13a,katori15a} and experiment~\cite{ushijima18a,dorscher22a,kim22a} is inconsistent even in the sign, as can be seen clearly from Fig~\ref{f1}. In 2018, Porsev {\em et al.} reported a value of $2.80(36)\times 10^{-5}$ a.u. ($\sim$ 0.339(44) mHz) by using the configuration interaction combined linearized coupled-cluster (CI+all-order) method~\cite{porsev18a}. Another result, $2.68(94)\times 10^{-5}$ a.u. ($\sim$ 0.324(15) mHz), was obtained by using the combined method of Dirac-Fock plus core polarization (DFCP) and relativistic configuration interaction (RCI) approaches~\cite{wu2019a}. Unexpectedly, both of these theoretical results have opposite signs to the measured value of $-0.962(40)$ mHz by RIKEN~\cite{ushijima18a}, despite of agreeing with each other. Recently, PTB and JILA reported independent experimental determinations of the E2-M1 polarizability difference of $-987^{+174}_{-223}$~\cite{dorscher22a} $\rm{\mu Hz}$ and $-1.24(5)$ mHz~\cite{kim22a}, respectively. Both experimental results have the same negative sign as the measurement by RIKEN. The inconsistency between theory and experiment sharpens.  

Since the ratio of E2/E1 polarizabilities is of the order of $\alpha Z_{core}S$($Z_{core}$ is the core charge, $S$ is the quadrupole shape factor), and the ratio of M1/E1 is of the order of $\alpha Z_{core}$, to calculate E2 and M1 polarizabilities, relativistic formalism is needed. When using the sum-over-states method to calculate the multipolar E2 and M1 polarizabilities, it is crucial to keep the completeness of intermediates states. Therefore, we need to include the virtual electron-positron pair contribution in the intermediate states, i.e., the Dirac negative-energy-states (hole, virtual positron) contribution. The importance of this has been emphasized in the calculations of $g$-factor of atoms and ions~\cite{shabaev02a,lindroth93a,glazov04a,wagner13a,agababaev18a,arapoglou19a,cakir20a,wu22a}. However, the contribution of negative-energy states to the multipolar polarizabilities for the optical clocks has never been discussed before.

In the present work, we take account of the negative-energy-states contributions to the dynamic multipolar polarizabilities using improved DFCP+RCI method. We find that for the M1 polarizability, the negative-energy-states contribution is much larger than that of positive-energy states by several orders of magnitude. For the Sr clock, the E2-M1 polarizability difference is determined to be $-7.74(3.92)\times10^{-5}$ a.u. [-0.935(477) mHz], agreed with all the experimental results. Our work has eliminated the sign inconsistency for E2-M1 polarizability difference between theory and experiment, and confirms the importance of negative-energy states on the M1 polarizability for optical clocks.

\begin{figure}
	\includegraphics[width=0.5\textwidth]{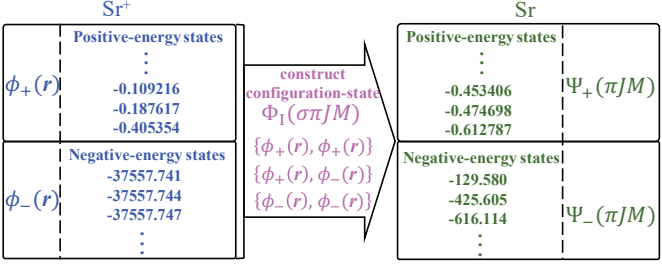}
	\caption{\label{f2}(Color online) State diagrams for obtaining the positive- and negative-energy states of the Sr clock. $\phi_+(\bm{r})$ and $\phi_-(\bm{r})$ are the two branches of the Sr$^+$ wavefunctions, which are used to construct configuration state wavefunctions $\Phi_I(\sigma{\pi} JM)$. The $\Psi_+(\pi JM)$ and $\Psi_-(\pi JM)$ are the two branches of the Sr wavefunctions.}
\end{figure}

\emph{Theoretical Method.} Different from available calculations, all the positive-energy and negative-energy states of monovalent-electron ion are used to construct the configurations of divalent electron atom. The summation in the formula of multipolar polarizabilities involves all the negative-energy and positive-energy states. Fig.~\ref{f2} shows the relation of positive-energy states and negative-energy states involved for the Sr clock. The detailed implementation follows: 

First, the core-orbital wavefunctions $\psi(\bm{r})$ of frozen core are obtained by the Dirac-Fock (DF) calculation~\cite{tang13b}, which is used to construct the DF potential $V_{DF}(\bm{r})$ between a valence-electron and the core for utilization in subsequent calculations.

Second, the monovalent-electron wavefunctions, which include two branches  $\phi_+(\bm{r})$ and $\phi_-(\bm{r})$, corresponding to the wavefunctions of positive-energy and negative-energy states of the monovalent-electron ion, can be obtained by solving the following DFCP equation,  
\begin{equation}
	h_{\rm DFCP}(\bm{r})\phi_{\pm}(\bm{r})=\varepsilon\phi_{\pm}(\bm{r})
	\,,\label{e3}
\end{equation}
where $h_{\rm DFCP}(\bm{r})$ represents the DFCP Hamiltonian,
\begin{equation}
	h_{\rm DFCP}(\bm{r})=c{{\bm{\alpha}}}\cdot{\mathbf p}+(\beta-1)c^{2}-\frac{Z}{r}+V_{DF}(\bm{r})+V_{\rm 1}(\bm{r})
	\,,	\label{e4}
\end{equation}
with $\bm{\alpha}$ and $\beta$ the $4\times 4$ Dirac matrices, $\mathbf p$ the momentum operator, and $V_{\rm 1}(\bm{r})$ the one-body core-polarization potential~\cite{wu2019a,mitroy88c}.

For the monovalent-electron Mg$^+$, Ca$^+$, and Sr$^+$ ions studied in the present paper, we only need to perform these first two steps to obtain the basic structure information of the ions. But for the divalent-electron atoms, such as Mg, Ca, Sr, Cd, we also need to carry out the following configuration interaction calculations. 

Using the monovalent-electron ion wavefunctions $\phi_+(\bm{r})$ and $\phi_-(\bm{r})$ obtained in the second step, we can construct the configuration-state wavefunctions $\Phi_{I}(\sigma{\pi} JM)$, based on three different combinations of $\{\phi_+(\bm{r}),\phi_+(\bm{r})\}$, $\{\phi_+(\bm{r}),\phi_-(\bm{r})\}$, and $\{\phi_-(\bm{r}),\phi_-(\bm{r})\}$, to 
form a new configuration space for the calculations of divalent-electron atoms. The wavefunction of divalent-electron atoms can be obtained by solving the following eigen equation,
\begin{widetext}
\begin{eqnarray}
\bigg[\sum_{i=1}^{2}h_{\rm DFCP}(\bm{r}_i)+\frac{1}{\bm{r}_{12}}+V_{2}(\bm{r}_{12})\bigg]|\Psi_{\pm}({\pi}JM)\rangle=E|\Psi_{\pm}({\pi}JM)\rangle \,,\label{e7}
\end{eqnarray}
\end{widetext}
where $V_{2}(\bm{r}_{12})$ is two-body core-polarization interaction~\cite{mitroy10a,mitroy03f,wu2019a}.

The wavefunction $|\Psi_{\pm}({\pi}JM)\rangle$ with parity $\pi$, angular momentum $J$, and magnetic quantum number $M$ is also divided into two branches, the positive-energy states $|\Psi_+({\pi}JM)\rangle$ and the negative-energy states $|\Psi_{-}({\pi}JM)\rangle$, which can be expressed as a linear combination of the configuration-state wavefunctions, 
\begin{eqnarray}
	|\Psi_{\pm}({\pi}JM)\rangle=\sum_{I}{C_{I}|\Phi_{I}(\sigma{\pi} JM)\rangle}
	\,,\label{e10}
\end{eqnarray}
where $C_{I}$ and $\sigma$ respectively denote the expansion coefficients and the additional quantum number that serve to uniquely define each configuration state. 

\emph{E2-M1 polarizabilities.} We follow Ref.~\cite{wu2019a} to include the negative-energy states in our derivation and calculation. When an ion or atom exposed under a linear polarized laser field with the laser frequency $\omega$, the general expression of dynamic M1 polarizability $\alpha^{M1}(\omega)$ for the initial state $|n_0J_0M_0\rangle$ (where $n_0$ represents all other quantum numbers) is derived as
\begin{eqnarray}
	\alpha^{M1}(\omega)&=&\alpha^{M1}_S(\omega)+g_2(J_0,M_0)\alpha^{M1}_T(\omega)
	\,,\label{e11} 
\end{eqnarray}
\begin{eqnarray}
	\alpha^{M1}_S(\omega)=\frac{2}{3(2J_0+1)}\sum_{n\pm}\frac{\Delta E_{n0}|\langle n_0J_0\|T_{M1}\|nJ_n\rangle|^2}{\Delta E_{n0}^2-\omega^2}
	\,,\label{e12} 
\end{eqnarray}
\begin{eqnarray}
	\alpha^{M1}_T(\omega)&=&\sqrt{\frac{40J_0(2J_0-1)}{3(2J_0+3)(J_0+1)(2J_0+1)}}\sum_{n\pm}(-1)^{J_0+J_n}  \nonumber \\
	&\times&\begin{Bmatrix} 1&1&2\\J_0&J_0&J_n \end{Bmatrix}\frac{\Delta E_{n0}|\langle n_0J_0\|T_{M1}\|nJ_n\rangle|^2}{\Delta E_{n0}^2-\omega^2}
	\,,\label{e13} 
\end{eqnarray}
and
\begin{eqnarray}
	g_2(J_0,M_0)=\frac{3M_{0}^2-J_0(J_0+1)}{J_0(2J_0-1)}, \quad J_0 > \frac{1}{2}  
	\,,\label{e14}
\end{eqnarray}
with $\alpha^{M1}_S(\omega)$ and $\alpha^{M1}_T(\omega)$ are the scalar and tensor M1 polarizabilities, respectively. In Eqs.~(\ref{e12}) and (\ref{e13}), $T_{M1}$ is M1 transition operator, $\Delta E_{n0}$ is transition energy between initial state $|n_0J_0\rangle$ and intermediate state $|nJ_n\rangle$. For monovalent-electron ions, the summation index $n_{\pm}$ runs over all the positive-energy states $\phi_+(\bm{r})$ and negative-energy states $\phi_-(\bm{r})$ of the intermediate state. For divalent-electron atoms, the summation index $n_{\pm}$ runs over all the positive-energy states $\Psi_+(\bm{r})$ and negative-energy states $\Psi_-(\bm{r})$ of intermediate states. 

Similarly, using second-order perturbation theory, we can derive the general formula for dynamic E2 polarizability of the initial state $|n_0J_0M_0\rangle$,
\begin{eqnarray}
	\alpha^{E2}(\omega)&=&\frac{1}{30}(\alpha\omega)^2[\alpha^{E2}_S(\omega)+g_2(J_0,M_0)\alpha^{E2}_{T1}(\omega)\nonumber \\
	&+&g_4(J_0,M_0)\alpha^{E2}_{T2}(\omega)]
	\,,\label{e15}
\end{eqnarray} 
with $\alpha$ the fine structure constant, $\alpha^{E2}_S(\omega)$, $\alpha^{E2}_{T1}(\omega)$, and $\alpha^{E2}_{T2}(\omega)$ are the scalar and tensor E2 polarizabilities, derived as
\begin{eqnarray}
	\alpha^{E2}_S(\omega)=\frac{1}{(2J_0+1)}\sum_{n\pm}\frac{\Delta E_{n0}|\langle n_0J_0\|T_{E2}\|nJ_n\rangle|^2}{\Delta E_{n0}^2-\omega^2}
	\,,\label{e16} 
\end{eqnarray}
\begin{small}
\begin{eqnarray}
	\alpha^{E2}_{T1}(\omega)&=&5\sqrt{\frac{10J_0(2J_0-1)}{7(2J_0+3)(J_0+1)(2J_0+1)}}\sum_{n\pm}(-1)^{J_0+J_n+1}\nonumber \\
	&\times&\begin{Bmatrix} 2&2&2\\J_0&J_0&J_n \end{Bmatrix}\frac{\Delta E_{n0}|\langle n_0J_0\|T_{E2}\|nJ_n\rangle|^2}{\Delta E_{n0}^2-\omega^2}
	\,,\label{e17}
\end{eqnarray}
\end{small}
\begin{small}
\begin{eqnarray}
	\alpha^{E2}_{T2}(\omega)&=&9\sqrt{\frac{10J_0(J_0-1)(2J_0-1)(2J_0-3)}{7(2J_0+5)(2J_0+4)(2J_0+3)(2J_0+2)(2J_0+1)}}  \nonumber \\
	&\times&\!\!\!\!\sum_{n\pm}(-1)^{J_0+J_n}\!\begin{Bmatrix} 2&2&4\\J_0&J_0&J_n \end{Bmatrix}\!\frac{\Delta E_{n0}|\langle n_0J_0\|T_{E2}\|nJ_n\rangle|^2}{\Delta E_{n0}^2-\omega^2}
	\,,\label{e18} \nonumber \\
\end{eqnarray}
\end{small}
$T_{E2}$ in Eqs.~(\ref{e16})-(\ref{e18}) is the E2 transition operator. $g_4(J_0,M_0)$ in Eq.~(\ref{e15}) is 
\begin{small}
	\begin{eqnarray}
		g_4(J_0,M_0)&=&\frac{3(5M_0^2-J_0^2-2J_0)(5M_0^2+1-J_0^2)}{J_0(J_0-1)(2J_0-1)(2J_0-3)}\nonumber \\
		&-&\frac{10M_0^2(4M_0^2-1)}{J_0(J_0-1)(2J_0-1)(2J_0-3)}, \quad J_0 > \frac{3}{2}   
		\,.\label{e19} \nonumber \\
	\end{eqnarray}
\end{small}

The reduced matrix elements $\langle n_0J_0\|T_{M1}\|nJ_n\rangle$ and $\langle n_0J_0\|T_{E2}\|nJ_n\rangle$ can be expressed by the reduced matrix elements $\langle i\|t_{M1}\|k\rangle$ and $\langle i\|t_{E2}\|k\rangle$ of monovalent-electron system~\cite{johnson06a}, 
\begin{eqnarray}
	\langle i\|t_{M1}\|j\rangle&=&\frac{\kappa_i+\kappa_j}{2}\langle-\kappa_i\|C^1\|\kappa_j\rangle \nonumber \\
	&&\int r[P_i(r)Q_j(r)+Q_i(r)P_j(r)]dr
	\,,\label{e20} \\
	\langle i\|t_{E2}\|j\rangle&=&\langle\kappa_i\|C^2\|\kappa_j\rangle \nonumber \\
	&&\int r^2[P_i(r)P_j(r)+Q_i(r)Q_j(r)]dr
	\,,\label{e21}
\end{eqnarray}
where $P_i(r)$ and $Q_i(r)$ are the large and small components of wavefunctions for monovalent-electron system. Comparing Eq.(\ref{e20}) with Eq.(\ref{e21}), we can see that the radial integrations of M1 reduced matrix elements involves the cross product term of $P_i(r)$ and $Q_i(r)$, while the E2 reduced matrix elements do not contain them. 

\emph{Results and Discussions.} Using the improved DFCP+RCI method with negative-energy states included, we have performed comprehensive calculations of dynamic multipolar polarizabilities for the current developing clocks. We find that with inclusion of the negative-energy states, the effect of negative-energy states on E2 polarizability is weak and cannot be reflected under present theoretical accuracy, but the contributions of negative-energy states to M1 polarizability for all the clocks are dominant.

\begin{table}[!htbp]
\caption{\label{t1} Itemized contributions (Contr.) to the dynamic E2 polarizability (in a.u.) for the $5s^2\,^1S_0$ and $5s5p\,^3P_0^o$ states of the Sr clock at the 813.4280(5) nm magic wavelength. Tail represents the contribution from other positive-energy states, $\alpha^{E2+}$ and $\alpha^{E2-}$ represent the total contribution from positive-energy and negative-energy states, respectively. The numbers in the square brackets denote powers of ten.}
\begin{ruledtabular}
\begin{tabular}{llll}
\multicolumn{2}{c}{$5s^2\,^1S_0$}&\multicolumn{2}{c}{$5s5p\,^3P_0^o$}\\
\cline{1-2}\cline{3-4}
\multicolumn{1}{c}{Sub item} &\multicolumn{1}{c}{Contr.}
&\multicolumn{1}{c}{Sub item} &\multicolumn{1}{c}{Contr.}\\ \hline
$5s4d\,^3D_2$   &1.258[-7]     &$5s5p\,^3P_2^o$   &$-$2.805[-6]  \\
$5s4d\,^1D_2$   &6.965[-5]     &$5d5p\,^3F_2^o$   &3.095[-5]    \\
$5s5d\,^1D_2$   &1.224[-5]     &$5d5p\,^1D_2^o$   &3.149[-6]  \\
$5s5d\,^3D_2$   &1.106[-8]     &$5s6p\,^3P_2^o$   &1.741[-5]  \\
$5p^2\,^3P_2$   &5.966[-8]     &$4d5p\,^3D_2^o$   &3.603[-6]  \\
$5d^2\,^1D_2$   &3.887[-8]     &$5d5p\,^3P_2^o$   &2.139[-6]  \\
$5s6d\,^3D_2$   &4.981[-10]    &$5s4f\,^3F_2^o$   &2.644[-5]  \\
$5s6d\,^1D_2$   &1.226[-7]     &$5s7p\,^3P_2^o$   &2.601[-6]  \\
$5s7d\,^1D_2$   &2.600[-6]     &$5s5f\,^3F_2^o$   &8.768[-6]  \\
Tail            &7.950[-6]     &Tail              &3.214[-5]        \\
$\alpha^{E2+}$  &9.28[-5]      &$\alpha^{E2+}$    &12.44[-5]     \\
$\alpha^{E2-}$  &$-$8.64[-16]  &$\alpha^{E2-}$    &$-$1.10[-15]   \\
Total           &9.28[-5]      &Total             &12.44[-5]      \\
\end{tabular}
\end{ruledtabular}
\end{table}
\begin{table}[!htbp]
\caption{\label{t2} Itemized contributions (Contr.) to the dynamic M1 polarizability (in a.u.) for the $5s^2\,^1S_0$ and $5s5p\,^3P_0^o$ states of the Sr clock at the 813.4280(5) nm magic wavelength. Tail represents the contribution from other positive-energy states, $\alpha^{M1+}$ and $\alpha^{M1-}$ represent the total contribution from positive-energy and negative-energy states, respectively. The numbers in the square brackets denote powers of ten.}
\begin{ruledtabular}
\begin{tabular}{llll}
\multicolumn{2}{c}{$5s^2\,^1S_0$}&\multicolumn{2}{c}{$5s5p\,^3P_0^o$}\\
\cline{1-2}\cline{3-4}
\multicolumn{1}{c}{Sub item} &\multicolumn{1}{c}{Contr.}
&\multicolumn{1}{c}{Sub item} &\multicolumn{1}{c}{Contr.}\\ \hline
$5s4d\,^3D_1$    & 1.483[-15]      &$5s5p\,^3P_1^o$ &$-$4.811[-6] \\
$5s6s\,^3S_1$    & 4.098[-13]      &$5s5p\,^1P_1^o$ &$-$2.702[-7] \\
$5s5d\,^3D_1$    & 1.273[-12]      &$5s6p\,^3P_1^o$ &7.336[-10]    \\
$5p^2\,^3P_1$    & 1.539[-9]       &$5s6p\,^1P_1^o$ &1.766[-8]       \\
Tail             & 5.81[-10]       &Tail            &1.35[-8]     \\
$\alpha^{M1+}$   & 2.17[-9]        &$\alpha^{M1+}$  &$-$5.05[-6]   \\
$\alpha^{M1-}$   & $-$3.84[-4]     &$\alpha^{M1-}$  &$-$4.88[-4]   \\
Total            & $-$3.84[-4]     &Total           &$-$4.93[-4]   \\
\end{tabular}
\end{ruledtabular}
\end{table}
Tables~\ref{t1} and \ref{t2} list the itemized contributions to the dynamic E2 and M1 polarizabilities at the 813.4280(5) nm~\cite{ye09a} magic wavelength for the Sr clock, respectively. For E2 polarizability, the contribution of negative-energy states is less than $10^{-14}$ for both of the $5s^2\,^1S_0$ and $5s5p\,^3P_0^o$ clock states, and can be neglected. However, the contribution of negative-energy states dominates the dynamic M1 polarizability. For the $5s^2\,^1S_0$ state, with the negative-energy states, the dynamic M1 polarizability at the 813.4280(5) nm magic wavelength changes from $2.17\times 10^{-9}$ a.u. to $-3.84\times 10^{-4}$ a.u.  Similarly, for the $5s5p\,^3P_0^o$ state, the contribution of negative-energy states accounts for 99\% of the M1 polarizability. 

To investigate the key reason for the negative-energy-states contribution, we further analyze their individual contributions. We find that, unlike positive-energy states, the contribution from negative-energy states is not primarily from a few intermediate states, but rather from a cumulative effect of thousands of states with energies ranging from $-37559$ a.u. to $-37557$ a.u. ($2mc^2\approx37558$ a.u.). Although all of these negative-energy states with energies of $-37558(1)$ a.u. are far from the initial state, their radial wavefunctions $Q_j(r)$ have large overlap with $P_i(r)$ component of the initial state wavefunction, which results in the large $P_i(r)Q_j(r)$ product in Eq.~(\ref{e20}). In other words, it is a series of large M1 transition matrix elements between the negative-energy states and the initial state that lead to the dominant contribution of negative-energy states to the M1 polarizability.

\begin{table}[ht]
\caption{\label{t3} Summarized results of dynamic E2 and M1 polarizabilities (in a.u.) at the 813.4280(5) nm magic wavelength for the $5s^2\,^1S_0$ and $5s5p\,^3P_0^o$ states of the Sr clock. $\Delta\alpha^{E2}(\omega)$ and $\Delta\alpha^{M1}(\omega)$ are the dynamic E2 and M1 polarizability difference for the clock states, respectively. And $\Delta\alpha^{QM}(\omega)=\Delta\alpha^{M1}(\omega)+\Delta\alpha^{E2}(\omega)$. The numbers in parentheses are the theoretical and computational uncertainties. The numbers in the square brackets denote powers of ten.}
\begin{ruledtabular}
\begin{tabular}{lrrr}
\multicolumn{1}{l}{Polarizability}&\multicolumn{1}{c}{Present}&\multicolumn{1}{c}{Ref.~\cite{wu2019a}}
&\multicolumn{1}{c}{Ref.~\cite{porsev18a}} \\ \hline
  $\alpha^{E2}_{\,^1S_0}(\omega)$   &  9.28(57)[-5]       &9.26(56)[-5]     &8.87(26)[-5]  \\
\specialrule{0em}{1pt}{1pt}
  $\alpha^{E2}_{\,^3P_0^o}(\omega)$ &  12.44(76)[-5]      &12.44(76)[-5]    &12.2(25)[-5]  \\
\specialrule{0em}{1pt}{1pt}
  $\Delta\alpha^{E2}(\omega)$       &  3.16(95)[-5]       &3.18(94)[-5]     &3.31(36)[-5]  \\ \hline
\specialrule{0em}{1pt}{1pt}
  $\alpha^{M1}_{\,^1S_0}(\omega)$   & $-$3.84(24)[-4]     &2.12(13)[-9]     &2.37[-9]      \\
\specialrule{0em}{1pt}{1pt}
  $\alpha^{M1}_{\,^3P_0^o}(\omega)$ & $-$4.93(30)[-4]     &$-$5.05(31)[-6]  &$-$5.08[-6]   \\
\specialrule{0em}{1pt}{1pt}
  $\Delta\alpha^{M1}(\omega)$       & $-$1.09(38)[-4]     &$-$5.05(31)[-6]  &$-$5.08[-6]   \\  \hline
\specialrule{0em}{1pt}{1pt}
  $\Delta\alpha^{QM}(\omega)$       & $-$7.74(3.92)[-5]   &2.68(94)[-5]     &2.80(36)[-5]  \\
\end{tabular}
\end{ruledtabular}
\end{table}

Since the values of DFCP+RCI method for the E1 polarizability of the Sr, Mg, and Cd clocks agree with the results of CI+all-order method within 3\% ~\cite{wu2019a,wu2020a,zhou2021a}, we conservatively give 3\% error to all the reduced matrix elements for evaluating the uncertainty of present E2 and M1 polarizabilities. The results and a detailed comparison of the Sr clock are summarized in Table~\ref{t3}. Present E2 polarizability is in good agreement with the results reported in previous studies~\cite{porsev18a,wu2019a}, which only considered the contribution of positive-energy states. In contrast, the result of M1 polarizability $\Delta\alpha^{M1}(\omega)$ is two orders of magnitude larger than the values of Refs.~\cite{porsev18a,wu2019a}. 

Adding $\Delta\alpha^{M1}(\omega)$ and $\Delta\alpha^{E2}(\omega)$ together, we can obtain the E2-M1 polarizability difference $\Delta\alpha^{QM}(\omega)=-7.74(3.92)\times10^{-5}$ a.u. for the Sr clock, which includes the negative-energy-states contribution of $-1.04(38)\times 10^{-4}$ a.u. Compared with our previous value of $2.68(94)\times 10^{-5}$ a.u.~\cite{wu2019a}, the large uncertainty in present work is due to the dominant contribution of the differential M1 polarizability $\Delta\alpha^{M1}(\omega)$. Since the absolute value of $-1.09(38)\times 10^{-4}$ a.u. is an order of magnitude larger than the E2 polarizability difference $3.16(95)\times 10^{-5}$ a.u., the addition of two terms causes the cancellation of significant digits. 

To compare with experiments of the Sr clock directly, we need to convert all the theoretical values of $\Delta\alpha^{QM}(\omega)$ from a.u. to Hz by using the formula $\tilde{\alpha}^{QM}=\Delta \alpha^{QM}(\omega)E_R/\alpha^{E1}(\omega)$, where $\alpha^{E1}(\omega)=287(17)$ a.u. is the present dynamic E1 polarizability at 813.4280(5) nm~\cite{ye09a} magic wavelength, and $E_R$ is the lattice photon recoil energy~\cite{ushijima18a}. The comparison is plotted in Fig.~\ref{f1}. Our value of $-0.935(477)$ mHz with the negative-energy-states contribution is in good agreement with the three measured results of $-0.962(40)$~\cite{ushijima18a}, $-0.987^{+0.174}_{-0.223}$~\cite{dorscher22a} and $-1.24(5)$ mHz~\cite{kim22a}. This illustrates that the negative-energy states are crucial to multipolar polarizabilities of the Sr clock. In addition, there is a tension between the measurement of JILA~\cite{kim22a} and that of RIKEN~\cite{ushijima18a}. Therefore, development of high-accuracy theoretical methods with negative-energy states included is urgently needed to relax this tension.

\begin{figure}
	\includegraphics[width=0.54\textwidth]{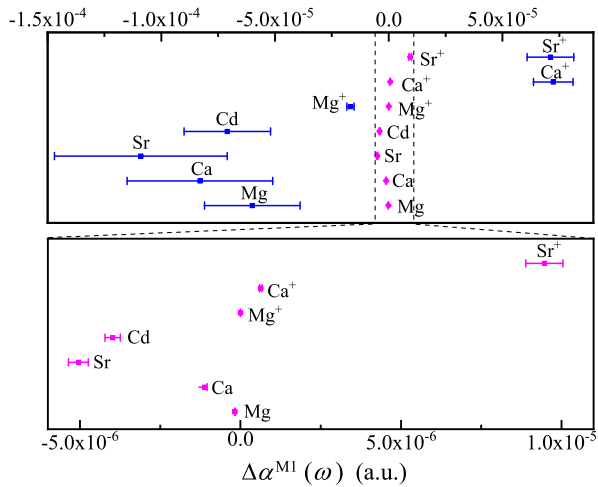}
	\caption{\label{f3}(Color online) Comparison of the M1 polarizability difference $\Delta\alpha^{M1}(\omega)$ (in a.u.) for current developing optical clocks. Values in magenta are the $\Delta\alpha^{M1}(\omega)$ without taking into account the negative-energy-states contribution, and values in blue are the $\Delta\alpha^{M1}(\omega)$ included both of the contributions from the positive-energy and negative-energy states.}
\end{figure}

Furthermore, we apply the present method to investigate the contribution of negative-energy states to the dynamic E2 and M1 polarizabilities of other optical clocks. All the results for the Mg, Ca, Cd, Mg$^+$, Ca$^+$, and Sr$^+$ clocks are summarized in the Supplemental Material~\cite{supplemental}. Similarly, for the E2 polarizability at the magic wavelengths, the negative-energy-states contribution is less than $10^{-15}$ a.u., which can be neglected. For the M1 polarizability, a concise comparison of M1 polarizability difference is shown in Fig.~\ref{f3}, the magenta and blue lines represent $\Delta\alpha^{M1}(\omega)$ without and with taking into account the negative-energy-states contribution, respectively. For each clock, the result in blue has a obvious deviation from the value in magenta, which qualitatively demonstrates the importance of negative-energy-states contribution. Taking the Ca$^+$ ion as an example, which is expected for achieving all-optical trapping utilizing the magic wavelength at far resonance~\cite{liu15a,huang22b}, the value of $\Delta\alpha^{M1}(\omega)$ is increased by two orders of magnitude after including the negative-energy-states contribution. The M1 polarizability differences in Fig.~\ref{f3} for various optical clocks have confirmed again that the negative-energy-states contribution to the magnetic polarizability is prevailing.

\emph{Conclusions.} Motivated to solve the obvious inconsistency in sign for the E2-M1 polarizability difference between existing theory and experiment in the Sr clock, we develop the combined DFCP+RCI method with inclusion of negative-energy states, and apply it to comprehensive calculations of dynamic M1 and E2 polarizabilities for the current developing clocks. Our result of E2-M1 polarizability difference for the Sr clock is $-$7.74(3.92)$\times$10$^{-5}$ a.u., which has the same sign with all the measured values. For other ion and atom clocks, the contribution of negative-energy states to the M1 polarizability is also crucial. Therefore, present work has resolved the sign inconsistency for the E2-M1 polarizability difference in the Sr clock. It has also revealed the importance of negative-energy states that are missing in all previous calculations for optical clocks, which will be helpful to be included in evaluating the multipolar interaction between light and matter in the field of precision measurement physics. 

\emph{Acknowledgments.} We thank Yong-Hui Zhang for helpful discussions on the negative-energy states, and thank J. Chen, K.-L. Gao, and Z.-C. Yan for reading our paper. This work was supported by the National Natural Science Foundation of China under Grant Nos. 12174402, 12274423, and 12004124, and by the Nature Science Foundation of Hubei Province Nos.2019CFA058 and 2022CFA013.


  \end{document}


\title{ Supplemental Material for 

Role of negative-energy states on the E2-M1 polarizability of optical clocks }

\author{Fang-Fei Wu, Ting-Yun Shi, Wei-Tou Ni, and Li-Yan Tang$^{\dag}$~\footnotetext{\dag Email Address: lytang@apm.ac.cn}}

\affiliation {State Key Laboratory of Magnetic Resonance and Atomic and Molecular Physics, Wuhan Institute of Physics and
Mathematics, Innovation Academy for Precision Measurement Science and Technology, Chinese Academy of Sciences, Wuhan 430071, People's Republic of China}

\date{\today}

\maketitle

The dynamic E2 and M1 polarizabilities for the Mg, Ca, Cd, Mg$^+$, Ca$^+$, and Sr$^+$ clocks are summarized in Tables~\ref{t1} and \ref{t2}, respectively. For atom clocks, we list the E2 and M1 polarizabilities at the measured magic wavelengths~\cite{kulosa15a,degenhardt04a,yamaguchi19a}.  
For ion clocks, we list E2 and M1 polarizabilities at the magic wavelengths that far away from resonance, since all-optical trapping of ions using these magic wavelengths can effectively suppress the micromotion and ac-Stark shift, and improve the frequency stability of clocks~\cite{liu15a,enderleina12a,huang22b}.

In Table~\ref{t1}, since the contribution of negative-energy states to dynamic E2 polarizability is less than $10^{-15}$ a.u., it is neglected and not listed in Table~\ref{t1}. But for the dynamic M1 polarizability at magic wavelength, it is seen from Table~\ref{t2}, for the lower $^1S_0$ or $^2S_{1/2}$ state of clocks that studied in the present work, the M1 polarizability is increased by 5-9 orders of magnitude with including the negative-energy states, and the sign has been changed. For the upper $^3P_0^o$ or $^2D_{5/2}$ state of clocks, the contribution of negative-energy states still dominates the M1 polarizability. Comparing the values of $\Delta\alpha^{M1+}(\omega)$, the M1 polarizability difference $\Delta\alpha^{M1\pm}(\omega)$ with negative-energy-states contribution has changed obviously in magnitude.

\begin{table*}[ht]
	\caption{\label{t1} The dynamic E2 polarizabilities (in a.u.) at the magic wavelengths $\lambda_m$ (in nm) of Mg, Ca, Cd, Mg$^+$, Ca$^+$, and Sr$^+$ clocks. $\Delta\alpha^{E2}(\omega)$ represents the dynamic E2 polarizability difference for the clock states. The contribution of negative-energy states is less than $10^{-15}$, which can be neglected in present calculations. The numbers in parentheses are the theoretical and computational uncertainties. The numbers in square brackets denote powers of ten. }
	\begin{ruledtabular}
		\begin{tabular}{lrrrr}
			System &$\lambda_m$(nm) &$\alpha^{E2}_{^1S_0}(\omega)$&$\alpha^{E2}_{^3P_0^o}(\omega)$ &$\Delta\alpha^{E2}(\omega)$\\ \hline
			Mg     &468.46(21)$^a$  &4.25(26)[-5]  &1.02(6)[-4]  &5.95(65)[-5]\\
			Ca     &735.5(20)$^b$  &7.51(44)[-5]  &1.00(6)[-4] &2.49(48)[-5] \\
			Cd     &419.88(14)$^c$  &2.53(15)[-5]  &9.91(60)[-5] &7.38(62)[-5]\\
			\hline
			&                &$\alpha^{E2}_{^2S_{1/2}}(\omega)$&$\alpha^{E2}_{^2D_{5/2}}(\omega)$&$\Delta\alpha^{E2}(\omega)$\\
			Mg$^+$ &737             &2.73(17)[-6]  &4.54(28)[-5] &4.27(28)[-5]    \\
			Ca$^+$ &1056.37(9)$^d$  &1.13(7)[-5]   &6.23(38)[-7] &$-$1.07(7)[-5]  \\
			Sr$^+$ &1898            &3.79(23)[-6]  &4.69(29)[-7] &$-$3.32(23)[-6] \\
		\end{tabular}
	\end{ruledtabular}
	$^{\rm a\,}$ Ref.~\cite{kulosa15a}\,, $^{\rm b\,}$ Ref.~\cite{degenhardt04a}\,,$^{\rm c\,}$ Ref.~\cite{yamaguchi19a}\,,
	$^{\rm d\,}$ Ref.~\cite{huang22b}
\end{table*}
\begin{table*}[ht]
\caption{\label{t2} The dynamic M1 polarizabilities (in a.u.) at the magic wavelengths $\lambda_m$ (in nm) of the Mg, Ca, Cd, Mg$^+$, Ca$^+$, and Sr$^+$ clocks. $\alpha^{M1\pm}(\omega)$ and $\alpha^{M1+}(\omega)$ represent the dynamic M1 polarizability with and without the negative-energy-states contribution, respectively. $\Delta\alpha^{M1\pm}(\omega)$ and $\Delta\alpha^{M1+}(\omega)$  represent the dynamic M1 polarizability difference with and without the contribution of negative-energy states, respectively. The numbers in parentheses are the theoretical and computational uncertainties. The numbers in square brackets denote powers of ten. }
\begin{ruledtabular}
\begin{tabular}{lrrrlrrr}
  System &$\lambda_m$(nm) &$\alpha^{M1+}_{^1S_0}(\omega)$&$\alpha^{M1\pm}_{^1S_0}(\omega)$ &$\alpha^{M1+}_{^3P_0^o}(\omega)$&$\alpha^{M1\pm}_{^3P_0^o}(\omega)$&$\Delta\alpha^{M1+}(\omega)$ &$\Delta\alpha^{M1\pm}(\omega)$\\ \hline
Mg     &468.46(21)$^a$  &1.23(7)[-11] &$-$2.07(13)[-4]&$-$1.72(10)[-7]&$-$2.67(16)[-4]&$-$1.72(10)[-7]&$-$0.60(21)[-4] \\
Ca     &735.5(20)$^b$  &4.69(29)[-10]&$-$3.30(20)[-4]&$-$1.11(7)[-6] &$-$4.13(25)[-4]&$-$1.11(7)[-6] &$-$0.83(32)[-4]\\
Cd     &419.88(14)$^c$  &1.45(9)[-9]  &$-$1.82(11)[-4]&$-$3.98(24)[-6]&$-$2.53(15)[-4]&$-$3.98(24)[-6]&$-$0.71(19)[-4]\\
\hline
&                &$\alpha^{M1+}_{^2S_{1/2}}(\omega)$&$\alpha^{M1\pm}_{^2S_{1/2}}(\omega)$&$\alpha^{M1+}_{^2D_{5/2}}(\omega)$&$\alpha^{M1\pm}_{^2D_{5/2}}(\omega)$&$\Delta\alpha^{M1+}(\omega)$&$\Delta\alpha^{M1\pm}(\omega)$\\
Mg$^+$ &737             &2.15(13)[-13]&$-$7.95(48)[-5]&$-$4.35(26)[-9]&$-$2.47(15)[-4]&$-$4.35(26)[-9]&$-$1.68(16)[-5]\\
Ca$^+$ &1056.37(9)$^d$  &2.75(17)[-13]&$-$1.30(8)[-4] &6.33(39)[-7]   &$-$5.77(35)[-5]&6.33(39)[-7]   &7.23(87)[-5]\\
Sr$^+$ &1898            &4.64(28)[-13]&$-$1.52(9)[-4] &9.47(58)[-6]   &$-$8.09(49)[-5]&9.47(58)[-6]   &7.11(1.03)[-5]\\
\end{tabular}
\end{ruledtabular}
$^{\rm a\,}$ Ref.~\cite{kulosa15a}\,, $^{\rm b\,}$ Ref.~\cite{degenhardt04a}\,,$^{\rm c\,}$ Ref.~\cite{yamaguchi19a}\,,
$^{\rm d\,}$ Ref.~\cite{huang22b}
\end{table*}